\documentclass[aps,prd,onecolumn,nofootinbib]{revtex4-2} 
\usepackage{titlesec}
\titleformat*{\section}{\LARGE\bfseries}
\titleformat*{\subsection}{\Large\bfseries}
\titleformat*{\subsubsection}{\large\bfseries}

\usepackage{epsfig}
\usepackage{enumitem}

\usepackage{graphics}     
\usepackage{graphicx}     
\usepackage{subfigure}
\usepackage{longtable}     
\usepackage{url}          
\usepackage{bm}           
\usepackage{natbib}
\usepackage{textpos}
\usepackage{amsfonts,amsmath,amssymb,mathrsfs,bm,amsthm}
\usepackage{float}
\usepackage{booktabs}
\usepackage{array}
\usepackage{tabu}
\usepackage{dcolumn}
\usepackage{rotating}
\usepackage{ulem}

\newcommand{\RN}[1]{%
  \textup{\uppercase\expandafter{\romannumeral#1}}%
}

\usepackage{nicefrac}
\usepackage{amsthm}
\usepackage{color}
\usepackage{cancel}
\usepackage{appendix}
\usepackage{makecell}
\usepackage{upgreek}

\usepackage[colorlinks=true,linkcolor=black,citecolor=blue,urlcolor=blue,bookmarksopen]{hyperref}

\newcommand{\appropto}{\mathrel{\vcenter{
  \offinterlineskip\halign{\hfil$##$\cr
    \propto\cr\noalign{\kern2pt}\sim\cr\noalign{\kern-2pt}}}}}

\begin{document}


\title{\Large{
Comment on ``Quantum sensor networks as exotic field telescopes for multi-messenger astronomy''
}}


\author{{\large\bf \rule[30pt]{0pt}{0pt}
Yevgeny~V.~Stadnik 
}}


\affiliation{{\large \rule[25pt]{0pt}{0pt}
Kavli Institute for the Physics and Mathematics of the Universe (WPI), The University of Tokyo Institutes for Advanced Study, The University of Tokyo, Kashiwa, Chiba 277-8583, Japan
}}

\raggedbottom



\maketitle






\large

\textbf{
In the recent work \cite{Dailey:2021multimessenger}, it was claimed that networks of quantum sensors can be used as sensitive multi-messenger probes of astrophysical phenomena that produce intense bursts of relativistic bosonic waves which interact non-gravitationally with ordinary matter.  
The most promising possibility considered in \cite{Dailey:2021multimessenger} involved clock-based searches for quadratic scalar-type interactions, with greatly diminished reach in the case of magnetometer-based searches for derivative-pseudoscalar-type interactions and clock-based searches for linear scalar-type interactions.  
In this note, we point out that the aforementioned work overlooked the ``back action'' of ordinary matter on scalar waves with quadratic interactions and that accounting for back-action effects can drastically affect the detection prospects of clock networks.  
In particular, back action can cause strong screening of scalar waves near Earth's surface and by the apparatus itself, rendering clock experiments insensitive to extraterrestrial sources of relativistic scalar waves.  
Additionally, back-action effects can retard the propagation of scalar waves through the interstellar and intergalactic media, significantly delaying the arrival of scalar waves at Earth compared to their gravitational-wave counterparts and thereby preventing multi-messenger astronomy on human timescales.  
}

A real scalar field $\phi$ may interact non-gravitationally with standard-model fields via the following $\phi^2$ interactions: 
\begin{equation}
\label{quadratic_interactions}
\mathcal{L} = \frac{\phi^2}{(\Lambda'_\gamma)^2} \frac{F_{\mu \nu} F^{\mu \nu}}{4} - \sum_\psi \frac{\phi^2}{(\Lambda'_\psi)^2} m_\psi \bar{\psi} \psi  \, ,  
\end{equation}
where the first term represents the interaction of the scalar field with the electromagnetic field tensor $F$, while the second term represents the interaction of the scalar field with the standard-model fermion fields $\psi$, with $m_\psi$ being the ``standard'' mass of the fermion and $\bar{\psi} = \psi^\dagger \gamma^0$ the Dirac adjoint.\footnote{ \normalsize  We employ the natural system of units $\hbar = c =1$ throughout, where $\hbar$ is the reduced Planck constant and $c$ is the speed of light in vacuum.}  
The interaction parameters $\Lambda'_{\gamma,\psi}$ have dimensions of energy.  
Note that the choice of signs in Eq.~(\ref{quadratic_interactions}) is identical to those used in Ref.~\cite{Dailey:2021multimessenger}.  
In the presence of ordinary matter, the effective potential experienced by the scalar field differs from the bare potential $V (\phi) = m^2 \phi^2 / 2$ assumed in \cite{Dailey:2021multimessenger}, where $m$ is the bare scalar mass, and instead reads: 
\begin{equation}
\label{effective_scalar_potential}
V_\textrm{eff} (\phi) = \frac{m^2 \phi^2}{2} + \sum_{X = \gamma, e, N} \frac{\rho_X \phi^2}{\left( \Lambda'_X \right)^2}  \, ,  
\end{equation}
where we have assumed that the ordinary-matter fields are associated with nonrelativistic atoms.  
Here, $\rho_\gamma = |\boldsymbol{E}|^2 / 2 \approx - F_{\mu \nu} F^{\mu \nu} / 4$ is the Coulomb energy density of a nonrelativistic nucleus, and $\rho_e$ and $\rho_N$ are the electron and nucleon mass-energy densities, respectively.  
In the presence of ordinary matter, therefore, the effective scalar mass $m_\textrm{eff}$ depends on the local ordinary-matter density $\rho$: 
\begin{equation}
\label{effective_scalar_mass}
m_\textrm{eff}^2 (\rho) = m^2 + \sum_{X = \gamma, e, N} \frac{2 \rho_X }{\left( \Lambda'_X \right)^2}  \, .  
\end{equation}
We hence see that, in the presence of the $\phi^2$ interactions (\ref{quadratic_interactions}), ordinary matter ``back acts'' on the scalar field $\phi$.  
Analogous back-action effects in various models of scalar fields with $\phi^2$ interactions have previously been explored in different contexts; see, e.g., Refs.~\cite{Olive:2008screening,Khoury:2010screening,Burrage:2018screening,Hees:2018screening,Stadnik:2020screening} and references therein.

\begin{figure}[t!]
\centering
\includegraphics[width=13.0cm]{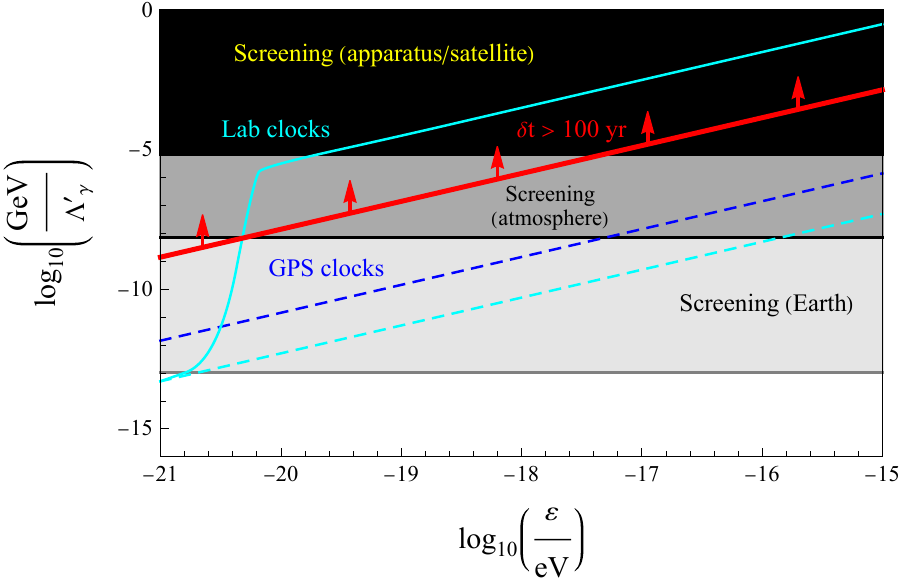}
\caption{ \normalsize  
Regions of parameter space for the quadratic interaction of a scalar field $\phi$ with the electromagnetic field (photon), as defined in Eq.~(\ref{quadratic_interactions}), where the scalar field is strongly screened near the surface of and inside Earth (light grey region), strongly screened by Earth's atmosphere (dark grey region), and strongly screened by a typical apparatus or satellite (black region).  
The solid and dashed cyan lines denote the estimated sensitivities of ground-based experiments using state-of-the-art optical clocks, with and without account of screening by Earth's non-gaseous matter, respectively.  
The dashed blue line denotes the estimated sensitivity of space-based experiments using microwave clocks.  
For the clock-based experiments, we have chosen burst parameters that yield a coherent burst over the entire relevant range of scalar particle energies $\varepsilon$, while matching the dashed cyan and blue lines to their counterparts in Fig.~3 of Ref.~\cite{Dailey:2021multimessenger}.  
The region above the solid red line denotes the region of parameter space where the time delay of a scalar wave relative to its gravitational-wave counterpart exceeds 100 years due to the back-action effects of the interstellar and intergalactic media on the scalar wave.  
See the main text and Methods section for more details.  
}
\label{Fig:Results_graph}
\end{figure}

The increase in the effective scalar mass in the presence of matter, according to Eq.~(\ref{effective_scalar_mass}), causes the scalar field to tend to be expelled from dense regions of matter, similarly to the expulsion of magnetic fields from the interior of a superconductor due to the generation of an effective photon mass inside the superconductor \cite{LL9:Stat_physics}.  
Consider a relativistic scalar wave with particle energy $\varepsilon$, propagating through vacuum and incident onto a dense spherical body of radius $R$.  
Strong screening of the scalar wave near the surface of and inside the dense body occurs when $m_\textrm{eff}^2 > \varepsilon^2$ inside the dense body and $q R \gtrsim 1$, where $q = \sqrt{m_\textrm{eff}^2 - \varepsilon^2}$.  
In Fig.~\ref{Fig:Results_graph}, we show the regions of parameter space for the scalar-photon interaction in Eq.~(\ref{quadratic_interactions}) where a scalar wave is strongly screened near the surface of and inside Earth, by Earth's atmosphere, and by a typical apparatus or satellite (see Methods \ref{Sec:Methods_Composition} for details of the assumed elemental compositions of these systems).  
The analysis presented in Ref.~\cite{Dailey:2021multimessenger} focused on the limiting case of small wave dispersion, corresponding to a coherent burst.\footnote{ \normalsize  We note that there is an inconsistency between the analysis presented in the text of Ref.~\cite{Dailey:2021multimessenger} and its implementation in Fig.~3 of \cite{Dailey:2021multimessenger}.  The analysis in Ref.~\cite{Dailey:2021multimessenger} explicitly assumed a small spread in the scalar particle energy, $\Delta \varepsilon \ll \varepsilon$, corresponding to a \textit{coherent} burst.  However, the burst duration of $\tau = 100~\textrm{s}$ assumed in Fig.~3 of \cite{Dailey:2021multimessenger} implies, by the time-energy uncertainty relation, a minimum particle energy spread of $\Delta \varepsilon \sim 1/ \tau \sim 10^{-16}~\textrm{eV}$.  This in turn implies that $\Delta \varepsilon \gtrsim \varepsilon$, which corresponds to an \textit{incoherent} burst, for the vast majority of scalar particle energies in Fig.~3 of \cite{Dailey:2021multimessenger}.  In our present note, we restrict ourselves to the simpler case of a coherent burst.}  
The relevant scattering problem can be solved analytically in the simple limiting case of a monochromatic continuous burst (see Methods \ref{Sec:Methods_Scattering_soln}).  
In the low-energy limit $pR \ll 1$, where $p = \sqrt{\varepsilon^2 - m^2}$ is the scalar particle momentum in vacuum, and when $qR \gg 1$, the scalar-wave amplitude at height $h \ll R$ above the surface of the dense body is suppressed by the factor $\approx h/R + 1/(qR)$; 
in the limit $q \to \infty$, one recovers the well-known result for hard-sphere scattering \cite{Griffiths:QM_book}.  
The leading anisotropic correction to the scalar-wave amplitude near the surface of the dense body is dipolar in nature and is suppressed compared to the dominant monopolar term by a factor of $\mathcal{O} (pR)$.  
Inside the dense body (or a dense spherical shell), the scalar-wave amplitude becomes exponentially suppressed at depths $d \gtrsim 1/q$.  
In Fig.~\ref{Fig:Results_graph}, we show how the sensitivity of ground-based clocks to the scalar-photon interaction is degraded when the effect of screening of a scalar wave by Earth's non-gaseous matter is taken into account, assuming a typical clock height of $h \sim 1~\textrm{m}$ above Earth's surface.  
It is apparent that, when the effects of screening of a scalar wave by Earth's non-gaseous matter and Earth's atmosphere are collectively taken into account, ground-based clocks become insensitive to extraterrestrial scalar waves when $\varepsilon \gtrsim 5 \times 10^{-21}~\textrm{eV}$ or $\Lambda'_\gamma \lesssim 10^7 - 10^8~\textrm{GeV}$, while space-based clocks become insensitive to scalar waves due to screening by the apparatus and satellite when $\Lambda'_\gamma \lesssim 10^4 - 10^5~\textrm{GeV}$.  
We hence see that space-based clock experiments can provide a more powerful probe of relativistic scalar waves than ground-based clock experiments, even when the space-based clocks are intrinsically less precise.  
Note the sharp transition from the weak screening regime to the strong screening regime for ground-based clocks (solid cyan line) around $\varepsilon \sim 10^{-21}~\textrm{eV}$ in Fig.~\ref{Fig:Results_graph}.

The increase in the effective scalar mass in the presence of matter, according to Eq.~(\ref{effective_scalar_mass}), also slows the propagation of a scalar wave through matter.  
In the limiting case when the scalar wave remains relativistic, the speed of the scalar wave is given by: 
\begin{equation}
\label{modified_speed}
v \approx 1 - \frac{m_\textrm{eff}^2}{2 \varepsilon^2}  \, ,  
\end{equation}
and the time delay between the arrival of a relativistic scalar wave and its gravitational-wave counterpart (assumed to be travelling at the speed of light) reads: 
\begin{equation}
\label{time-delay}
\delta t \approx \frac{m_\textrm{eff}^2 L}{2 \varepsilon^2}  \, ,  
\end{equation}
where $L$ is the distance from the source to detector, and we have assumed that the intervening medium is homogeneous.  
In Fig.~\ref{Fig:Results_graph}, we show the region of parameter space for the scalar-photon interaction in Eq.~(\ref{quadratic_interactions}) where the time delay exceeds 100 years due to the back-action effects of the interstellar and intergalactic media on the scalar wave.  
We have assumed the favourable scenario when $m_\textrm{eff}^2 \gg m^2$; on the other hand, if $m_\textrm{eff}^2 \approx m^2$, then $\delta t > 100~\textrm{years}$ is possible even in the limit $\Lambda'_\gamma \to \infty$.  
We have also assumed a source located outside of our Galaxy and that the scalar wave favourably enters our Galaxy parallel to the normal of the thin Galactic disk; in this case, the interstellar medium within our Galaxy gives the larger contribution to the time-delay effect, with a comparable contribution from the intergalactic medium only for sources located in the farthest reaches of the observable Universe.  
If the scalar wave instead propagates along a substantial fraction of the plane of the Galactic disk, e.g., if the source is located near the Galactic Centre, then the time-delay effect increases by at least an order of magnitude.  
We refer the reader to Methods \ref{Sec:Methods_Composition} for further details.  
We thus see that $\delta t > 100~\textrm{years}$ can occur in large regions of parameter space, which would prevent multi-messenger astronomy with clock-based experiments on human timescales and greatly exceed the figure $\delta t \lesssim 10~\textrm{hours}$ quoted in Ref.~\cite{Dailey:2021multimessenger} for the extragalactic source GW170608.  
Additionally, we note that depending on the mechanism responsible for producing relativistic scalar waves, the generation of an effective scalar mass in dense environments may strongly suppress the production of scalar waves in burst-type phenomena if $m_\textrm{eff}^2 > \varepsilon^2$ at the source.

In this note, we have highlighted how back-action effects can drastically affect the detection prospects of clock networks in searches for relativistic scalar waves with the quadratic interactions in Eq.~(\ref{quadratic_interactions}).  
For $\phi^2$ interactions of the type in Eq.~(\ref{quadratic_interactions}) but with the signs reversed, the scalar field will tend to be anti-screened by dense regions of matter when $0 < m_\textrm{eff}^2 < m^2$.  
However, for the clock-based sensitivity curve parameters in Fig.~3 of \cite{Dailey:2021multimessenger}, $m_\textrm{eff}^2 < 0$ for the matter densities encountered in Earth's interior and atmosphere, as well as apparata and satellite components;  
in this case, the effective potential has an unstable maximum at $\phi = 0$ and the classical scalar-field equation of motion admits a tachyonic solution, at least when higher-order terms in the bare potential, higher-dimensional operators and finite spatial extent of the dense region are neglected.  
For further discussion of such types of ``opposite-sign'' interactions, including discussion of possible non-perturbative effects, we refer the reader to Refs.~\cite{Hees:2018screening,Stadnik:2020screening,Damour:1993scalarisation,Suyama:2015scalarisation,Pretorius:2016scalarisation,Popchev:2016scalarisation}.  
In the case of linear scalar-type interactions, the interaction generates a source term in the classical equation of motion for the scalar field and hence does not give rise to back-action effects for a harmonic bare potential, when the classical scalar-field equation of motion is linear;  however, back-action effects can arise for more complicated potentials that give rise to non-linear term(s) in the scalar-field equation of motion (see, e.g., \cite{Burrage:2018screening,Stadnik:2020screening}).  
The back-action effects due to spin-unpolarised matter considered in our present note do not apply to derivative-pseudoscalar-type interactions at leading order.  
Possible back-action effects due to spin-\textit{polarised} matter, such as spin-polarised electrons in certain types of magnetic shielding and spin-polarised geoelectrons inside Earth, warrant separate investigation.

\vspace{200mm}

\appendix
\section{Methods}
\label{Sec:Methods}

\subsection{Elemental compositions and mass-energy contributions of matter}
\label{Sec:Methods_Composition}

The main mass-energy contributions in an electrically neutral atom containing $A$ nucleons and $Z$ electrons are as follows: 
\begin{equation}
\label{mass-energy_contributions_atom}
M_\textrm{atom} \approx A m_N + Z m_e + \frac{a_C Z (Z-1)}{A^{1/3}} + Z a_p + (A-Z) a_n  \, .  
\end{equation}
The first two terms in Eq.~(\ref{mass-energy_contributions_atom}) correspond to the nucleon and electron mass-energies, respectively.  
The third term corresponds to the energy associated with the electrostatic repulsion between protons in a spherical nucleus of uniform electric-charge density, with the coefficient $a_C \approx 3\alpha / (5 r_0) \approx 0.7~\textrm{MeV}$, where $\alpha \approx 1 / 137$ is the electromagnetic fine-structure constant and $r_0 \approx 1.2~\textrm{fm}$ is the internucleon separation parameter that is determined chiefly by the strong nuclear force.  
The final two terms in Eq.~(\ref{mass-energy_contributions_atom}) correspond to the electromagnetic energies of the proton and neutron, respectively, with the coefficients $a_p \approx + 0.63~\textrm{MeV}$ and $a_n \approx - 0.13~\textrm{MeV}$ derived from the application of the Cottingham formula \cite{Cottingham:1963n-p} to electron-proton scattering \cite{Gasser:1982quarks}.  
We apply Eq.~(\ref{mass-energy_contributions_atom}) to determine the fractional mass-energy contributions due to the electromagnetic, electron-mass and nucleon-mass components in various systems summarised in Table~\ref{tab:frac_mass-energies}, assuming the elemental compositions outlined below for these systems.

\begin{table}[h!]
\centering
\caption{\normalsize  
Summary of the fractional mass-energy contributions $K_X$ due to electromagnetic ($X = \gamma$), electron-mass ($X = e$) and nucleon-mass ($X = N$) components in various systems.  
}
\label{tab:frac_mass-energies}
\vspace{5mm}
\large
\begin{tabular}{ |c|c|c|c| }%
\hline
 System & $K_\gamma$ & $K_e$ & $K_N$  \\ \hline 
 Earth's non-gaseous interior & $1.9 \times 10^{-3}$ & $2.4 \times 10^{-4}$ & $1.0$  \\ \hline 
 Earth's atmosphere & $9.5 \times 10^{-4}$ & $2.7 \times 10^{-4}$ & $1.0$  \\ \hline 
 Interplanetary and interstellar media & $6.3 \times 10^{-4}$ & $4.4 \times 10^{-4}$ & $1.0$  \\ \hline 
\multicolumn{1}{c}{} & \multicolumn{1}{c}{} & \multicolumn{1}{c}{} & \multicolumn{1}{c}{}  \\ 
\end{tabular}
\end{table}

\textbf{Earth's non-gaseous interior ---}  
We assume that the elemental composition of Earth's non-gaseous interior is a $1:1:1$ ratio of $^{24}$Mg$^{16}$O, $^{28}$Si$^{16}$O$_{2}$ and $^{56}$Fe by number.  
We treat Earth as a uniform sphere of radius $R_\oplus \approx 6400~\textrm{km}$ and with a density of $\rho_\oplus \approx 5.5~\textrm{g/cm}^3$.

\textbf{Earth's atmosphere ---}  
We treat Earth's atmosphere as a $4:1$ ratio of $^{14}$N$_{2}$ and $^{16}$O$_{2}$ by number, with a constant density of $\rho \approx 10^{-3}~\textrm{g/cm}^3$ and extending out from Earth's surface to an altitude of $h \approx 10~\textrm{km}$.

\textbf{Apparata and satellites ---}  
The details of the apparatus, including the sizes, materials and geometries of the apparatus components, as well as the details of the surrounding shielding and the laboratory or satellite environment, vary between individual experiments.  
Atomic clocks consist of low-density atomic vapours or individual ions contained within a spherical or cylindrical vacuum chamber.  
Common materials for vacuum chamber walls include aluminium (density $\approx 3~\textrm{g/cm}^3$) and stainless steel (density $\approx 8~\textrm{g/cm}^3$).  
Ground-based clocks in laboratories are surrounded by reinforcing structures of buildings which are commonly made of reinforced concrete (density $\approx 2 - 3~\textrm{g/cm}^3$).  
For simplicity, we model a typical apparatus or satellite as a uniform sphere of radius $R \sim 10~\textrm{cm}$, with a density comparable to Earth's average density and assuming the same elemental composition quoted above for Earth's non-gaseous interior.

\textbf{Interstellar and intergalactic media ---}  
We assume that the elemental composition of the interstellar and intergalactic media is a $3:1$ ratio of $^{1}$H and $^{4}$He by mass, and have neglected the effects of stellar nucleosynthesis.  
We treat the interstellar and intergalactic media as homogeneous media, with particle densities of $\sim 10~\textrm{cm}^{-3}$ and $\sim 10^{-7}~\textrm{cm}^{-3}$, respectively.  
Our Galactic disk has a radius of $\approx 50~\textrm{klyr}$ and a thickness of $\sim 1~\textrm{klyr}$.  
The distance from the Sun to the Galactic Centre is $\approx 25~\textrm{klyr}$.  
The present-day size of the observable Universe is $\approx 100~\textrm{Glyr}$, while the distances from Earth to the recently observed extragalactic sources GW170608 and GW170817 are $\approx 1~\textrm{Glyr}$ and $\approx 100~\textrm{Mlyr}$, respectively.

\subsection{Scattering of a relativistic scalar wave by a dense spherical body}
\label{Sec:Methods_Scattering_soln}
Here, we solve the problem of a relativistic scalar wave, propagating through vacuum and incident onto a dense spherical body of radius $R$, when the scalar field has the effective potential (\ref{effective_scalar_potential}).  
We shall find it convenient to work with the complex scalar field $\Phi = (\phi + i \eta) / \sqrt{2}$, where $\phi$ and $\eta$ are two real scalar fields, and then take the real part of $\Phi$ at the end of the calculation.  
In this case, the differential equation for the complex scalar field reads: 
\begin{equation}
\label{master-differential-equation}
\frac{\partial^2 \Phi}{\partial t^2} - \boldsymbol{\nabla}^2 \Phi + \left[ m^2  + \frac{2 \rho_X (r)}{\left( \Lambda'_X \right)^2} \right] \Phi (t,r,\theta,\varphi) = 0  \,  ,  
\end{equation}
with an analogous equation for $\Phi^*$.  
In Eq.~(\ref{master-differential-equation}) and below, summation over the mass-energy components $X = \gamma, e, N$ is implicit.  
The radial density profile reads: 
\begin{equation}
\label{radial-density-profile}
\rho_X (r) = 
\left\{\begin{array}{ll}
\rho_X  \, , &  0\leq r < R  \, ; \\
0  \, , &  r > R \, . 
\end{array}\right.
\end{equation}
In the limit of a dispersionless scalar wave, $\Delta \varepsilon \to 0$, corresponding to a monochromatic continuous source that produces scalar particles with energy $\varepsilon$ in a non-expanding Universe, the solutions of the differential equation (\ref{master-differential-equation}) are stationary and can be factorised as follows: 
\begin{equation}
\label{DE-factorisation-soln}
\Phi (t,r,\theta,\varphi) = \sum_{l=0}^{\infty} \sum_{n=-l}^{+l} \chi_l (r) Y_l^{n} (\theta, \varphi) \exp(-i \varepsilon t)  \,  ,  
\end{equation}
where $Y_l^{n}$ are the spherical harmonic functions, and the radial functions $\chi_l$ are solutions of the following ordinary differential equation: 
\begin{equation}
\label{radial-ODE}
\frac{d^2 \chi_l}{d r^2} + \frac{2}{r} \frac{d \chi_l}{dr} + \left[ m_\textrm{eff}^2 [\rho (r)] - \varepsilon^2 - \frac{l (l+1)}{r^2}  \right] \chi_l (r) = 0  \,  .  
\end{equation}
Far away from the source and far away from the dense spherical body of interest, the generated scalar wave can be approximated by a plane wave, which we take to be travelling along the negative $z$-axis towards the dense spherical body, centred at the origin of the spatial coordinates.  
Since the dense body is spherically symmetric, the resulting scalar-field wave function must be independent of the azimuthal angle $\varphi$, and so only spherical harmonics with $n = 0$ can contribute to the scalar-field wave function in Eq.~(\ref{DE-factorisation-soln}).  
The external solution is the sum of the incident plane wave and reflected spherical waves.  
With the aid of the plane-wave expansion, the external solution that remains finite as $r \to \infty$ can be written in the following form: 
\begin{equation}
\label{external-solution-general-form}
\Phi_\textrm{out} (t,r,\theta) = A \sum_{l=0}^{\infty} (2l+1) i^l j_l (pr) P_l [\cos (\theta)] \exp(-i \varepsilon t) + \sum_{l=0}^{\infty} B_l h_l^{(1)} (pr) P_l [\cos (\theta)] \exp(-i \varepsilon t)  \,  ,  
\end{equation}
where $A$ and $B_l$ are parameters with dimensions of energy, $P_l$ is the Legendre polynomial of order $l$, $j_l$ is the spherical Bessel function of the first kind and order $l$, $h_l^{(1)}$ is the spherical Hankel function of the first kind and order $l$, and $p = \sqrt{\varepsilon^2 - m^2}$ is the scalar particle momentum in vacuum.  
Strong screening of the scalar wave near the surface of and inside the dense body occurs when $m_\textrm{eff}^2 > \varepsilon^2$ inside the dense body and $q R \gtrsim 1$, where $q = \sqrt{m_\textrm{eff}^2 - \varepsilon^2}$.  
In this case, the internal solution, which remains finite as $r \to 0$, reads as follows: 
\begin{equation}
\label{internal-solution-general-form}
\Phi_\textrm{in} (t,r,\theta) = \sum_{l=0}^{\infty} C_l y_{-l-1} (-iqr) P_l [\cos (\theta)] \exp(-i \varepsilon t)  \,  ,  
\end{equation}
where $C_l$ are parameters with dimensions of energy, and $y_l$ is the spherical Bessel function of the second kind and order $l$.  
The requirement of continuity of $\Phi$ and $d \Phi / d r$ at $r = R$, together with the orthogonality of Legendre polynomials with different values of $l$, fixes the coefficients $B_l$ and $C_l$ in Eqs.~(\ref{external-solution-general-form}) and (\ref{internal-solution-general-form}) in terms of the parameter $A$: 
\begin{equation}
\label{Bl_coefficients}
B_l = \frac{A i^l (2l+1) \left\{ py_{-l-1}(-iqR) [ j_{l+1}(pR) - j_{l-1}(pR) ] - iq j_l(pR) [ y_{-l-2}(-iqR) - y_{-l}(-iqR) ] \right\}}{iqh_l^{(1)}(pR) [ y_{-l-2}(-iqR) - y_{-l}(-iqR) ] + py_{-l-1}(-iqR) [ h_{l-1}^{(1)}(pR) - h_{l+1}^{(1)}(pR) ] }  \,  ,  
\end{equation}
\begin{equation}
\label{Cl_coefficients}
C_l = \frac{2A i^l (2l+1)}{ pR^2 \left\{ qh_l^{(1)}(pR) [ y_{-l-2}(-iqR) - y_{-l}(-iqR) ] - ip y_{-l-1}(-iqR) [ h_{l-1}^{(1)}(pR) - h_{l+1}^{(1)}(pR) ] \right\} }  \,  .  
\end{equation}
In the low-energy limit $pR \ll 1$, and when $qR \gg 1$, the dominant contribution to the scalar-field wave function at small heights $h \ll R$ above the surface of the dense body is monopolar in nature ($l = 0$) and reads: 
\begin{equation}
\label{external-solution-monopolar-term}
\Phi_\textrm{out} (t,r) \approx A \left( \frac{h}{r} + \frac{1}{qr} \right) \exp(-i \varepsilon t)  \,  ,  
\end{equation}
while the leading anisotropic correction to the scalar-field wave function at small heights is dipolar in nature ($l = 1$) and reads: 
\begin{equation}
\label{external-solution-dipolar-term}
\delta \Phi_\textrm{out} (t,r,\theta) \approx 3iApR \left( \frac{hR}{r^2} + \frac{R}{qr^2} \right) \cos(\theta) \exp(-i \varepsilon t)  \,  .  
\end{equation}
In the limit $q \to \infty$, we recover the well-known result for hard-sphere scattering \cite{Griffiths:QM_book}.\footnote{Klein's paradox for a complex scalar field does not apply in our present problem.  On the one hand, a strongly repulsive electrostatic potential $V$ causes the scalar particle momentum to be real inside the potential barrier when $V > \varepsilon + m$.  On the other hand, when the effective scalar particle mass inside a dense body satisfies $m_\textrm{eff}^2 > \varepsilon^2$, the scalar particle momentum becomes imaginary, resulting in the attenuation of the scalar-wave amplitude inside the dense body.}  
Meanwhile, the internal solution reads as follows at leading order: 
\begin{equation}
\label{internal-solution-1}
\Phi_\textrm{in} (t,r) \approx 2A \exp (-qR) \exp (-i \varepsilon t) ~~\textrm{for}~ qr \ll 1  \,  ,  
\end{equation}
\begin{equation}
\label{internal-solution-2}
\Phi_\textrm{in} (t,r) \approx \frac{A \exp [-q(R-r)]}{qr} \exp (-i \varepsilon t) ~~\textrm{for}~ qr \gg 1  \,  .  
\end{equation}
We see that at depths $d \gtrsim 1/q$ inside the dense body, the scalar-wave amplitude becomes exponentially suppressed.  
Likewise, the scalar-wave amplitude is also exponentially suppressed at depths $d \gtrsim 1/q$ inside a dense spherical shell (such as Earth's atmosphere).  
The exponential suppression (rather than abrupt vanishing) of the scalar-wave amplitude inside dense bodies when $m_\textrm{eff}^2 > \varepsilon^2$ is a consequence of the position-momentum uncertainty relation.

\section*{Acknowledgements}
The author thanks Josh Eby, Satoshi Shirai and Volodymyr Takhistov for discussions. 
The author thanks Andrei Derevianko and Derek Jackson Kimball for feedback on the manuscript. 
This work was supported by the World Premier International Research Center Initiative (WPI), MEXT, Japan, and by the JSPS KAKENHI Grant Number JP20K14460.

\section*{Author contributions}
The author solely conceived the project, developed all parts of the methodology and wrote the paper.

\section*{Competing interests}
The author declares no competing interests.

\vspace{100mm}

\section*{Rebuttal to arXiv:2112.02653v1}
The preceding parts of my present article solely contain my unaltered v1 comment \cite{Stadnik:2021comment} regarding the paper \cite{Dailey:2021multimessenger}. 
Here, I present a brief rebuttal to several erroneous and grossly misleading claims in the recent reply \cite{Derevianko:2021reply}. 
From the outset, it should be noted that the reply in \cite{Derevianko:2021reply} features only three of the seven original authors of the paper \cite{Dailey:2021multimessenger} under scrutiny. 
This immediately raises questions about the credibility of the reply in \cite{Derevianko:2021reply} --- it is unusual for a reply not to be endorsed by a majority of the authors of the original paper, let alone by the full set of original authors.

\begin{itemize}

\item 
The opening statement in \cite{Derevianko:2021reply} deceptively rewords the final sentence of the introductory paragraph in my comment \cite{Stadnik:2021comment}. 
This establishes a false basis for the misleading narrative repeated throughout \cite{Derevianko:2021reply}. 
Quite frankly, this is dishonest conduct that undermines the integrity of scientific discourse. 
From the comparison of the relevant statements (reproduced below), it is clear that my comment did \textbf{NOT} make the ``blanket claim'' concocted by Derevianko \textit{et al}. 

\textbf{Statement in reply \cite{Derevianko:2021reply}:}  \textit{The comment by Stadnik [arXiv:2111.14351v1] claims that ``back-action'', i.e.~interaction of exotic low-mass fields (ELF) with ordinary matter, ``prevents the multi-messenger astronomy on human timescales.'' We strongly disagree with this blanket claim.} 

\textbf{Statement in my comment \cite{Stadnik:2021comment}:} \textit{Additionally, back-action effects can retard the propagation of scalar waves through the interstellar and intergalactic media, significantly delaying the arrival of scalar waves at Earth compared to their gravitational-wave counterparts and thereby preventing multi-messenger astronomy on human timescales.}

\item
The reply in \cite{Derevianko:2021reply} claims to present a ``\textit{counter-argument which demonstrates that there remains a large parameter space for detecting ELFs}'' by focusing on $\phi^2$ interactions of opposite sign to those in their original paper \cite{Dailey:2021multimessenger} and in Eq.~(\ref{quadratic_interactions}) of my comment \cite{Stadnik:2021comment}. 
However, the basis of this claim in \cite{Derevianko:2021reply} is incorrect. 
As pointed out in my comment \cite{Stadnik:2021comment}, in the case of such ``opposite-sign'' $\phi^2$ interactions, we have $m_\textrm{eff}^2 < 0$ in a terrestrial-density environment for the clock and burst parameters assumed in \cite{Dailey:2021multimessenger}. 
The case $m_\textrm{eff}^2 < 0$ has an unstable tachyonic solution, which shares similarities with the well-known tachyonic instability in the standard model that is responsible for the Higgs scalar field developing a non-zero vacuum expectation value (see, e.g., Ref.~\cite{Griffiths_book_EPP}). 
The rudimentary analysis in \cite{Derevianko:2021reply} is therefore NOT valid for the case when $m_\textrm{eff}^2 < 0$. 
Indeed, \cite{Derevianko:2021reply} notes that the approximation behind their Eq.~(3) breaks down when $m_\textrm{eff}^2 < 0$ ($m_\textrm{eff}^2 = m^2 + \beta$ in the notation of \cite{Derevianko:2021reply}). 
The other regime for opposite-sign $\phi^2$ interactions discussed in \cite{Derevianko:2021reply}, namely $0 < m_\textrm{eff}^2 < m^2$, is NOT applicable in a terrestrial-density environment for the clock and burst parameters assumed in \cite{Dailey:2021multimessenger}.

\item
The reply in \cite{Derevianko:2021reply} states that ``\textit{Stadnik erroneously claims that the choice of sign in his Eq.~(1) is identical to our paper}''. 
A thorough examination of their original paper \cite{Dailey:2021multimessenger}, however, fails to reveal any explicit mention of $\phi^2$ interactions with signs that differ from those in Eq.~(\ref{quadratic_interactions}) of my comment \cite{Stadnik:2021comment}. 
The signs of the $\phi^2$ interactions are explicitly defined in \cite{Dailey:2021multimessenger} between their Eqs.~(1) and (2) on page 2, where it is evident that the signs precisely match those in Eq.~(\ref{quadratic_interactions}) of my comment \cite{Stadnik:2021comment}. 
If one follows the flow of the exposition in \cite{Dailey:2021multimessenger}, then the $\Gamma$ parameters in their Eqs.~(58,59) are logically determined by comparison with the aforementioned interaction Lagrangians appearing between Eqs.~(1) and (2) in \cite{Dailey:2021multimessenger}, otherwise the signs of the interactions in Eqs.~(58,59) would formally be undefined. 
As a side note, the sign of the standard-model term $\mathcal{L}_\textrm{SM}^f$ in Eq.~(1) of \cite{Derevianko:2021reply} is incorrect.

\item
Finally, Derevianko \textit{et al}.'s suggestion of improper attribution of communication with the principal author (Derevianko) is outrageous and devoid of factual merit. 
Prior to the submission of my comment \cite{Stadnik:2021comment}, Derevianko requested that I remove his name from my acknowledgements. 
Nevertheless, there was a clear acknowledgement in v1 of my comment \cite{Stadnik:2021comment} thanking Andrei Derevianko and Derek Jackson Kimball for feedback on my manuscript.

\end{itemize}


\end{document}